\documentclass{pasj00}
\draft

\begin{document}
\SetRunningHead{Richmond}{Stars in the SDF}
\Received{}
\Accepted{}

\title{Properties of Stars in the Subaru Deep Field}

\author{Michael \textsc{Richmond} }
  
\affil{Physics Department, Rochester Institute of Technology, Rochester, NY 14623}
\email{mwrsps@rit.edu}


%


\maketitle

\begin{abstract}
We investigate the properties of objects in
the Subaru Deep Field (SDF), using
public catalogs constructed from images in
several optical passbands.
Using a small subset of objects most likely
to be stars, we construct a stellar locus
in three-dimensional color space.
We then compare the position of all objects
relative to this locus to create larger samples of
stars in the SDF with rough spectral types.
The number counts of stars defined in this way 
are consistent with
those of current models of the Galaxy.
\end{abstract}

\section{Introduction}

During the spring seasons of 2002 and 2003,
the Subaru telescope and
Suprime-Cam mosaic camera 
acquired very deep images
in several passbands of a region of
the sky near the North Galactic Pole
(\cite{Kashikawa2004}).
This Subaru Deep Field (SDF) 
combines a relatively wide area
(roughly one third of a square degree)
and very deep detection limit
(down to magnitude $28$ in $B$),
yielding hundreds of thousands
of sources with well measured
properties.
The SDF project team has released to
the astronomical community the 
full imaging dataset as well as a
rich set of catalogs,
making the SDF a resource which is sure
to be used by many groups working in 
different areas over the next decade.

The primary goals of the SDF project
depend on extragalactic sources:
using a combination of broadband 
and narrowband measurements,
it is possible to identify objects at
redshifts of $z \simeq 4 - 7$;
one can also probe the luminosity 
function of nearby galaxies to very low
levels.
In these studies, foreground stars
in the Milky Way are annoying contaminants
which must be discarded in order to count
galaxies properly.
We realized, however, that one can also
use the SDF catalogs to study the properties
of stars themselves.
There are very few regions of the sky 
which have been surveyed as deeply 
as the SDF, and none at such a high
galactic latitude:
the Hubble Deep Field North 
(\cite{Williams1996}), 
for example,
is at $b = 54$ degrees,
close to an ecliptic pole,
as are most other regions studied by 
satellite-borne instruments.
The SDF, at
$l = 37$ and $b = 83$ degrees, provides valuable
information on the stellar content of our galaxy
in a relatively unexplored direction.

The aim of this paper is twofold:
first, 
to describe in some detail how one can 
use colors to isolate stars from galaxies
in deep imaging surveys;
second, to make available to the community
the result of our methods applied to 
sources in the SDF,
so that others may easily select new subsets
of stellar (or non-stellar) sources.
In Section 2,
we will describe how one use the 
colors of stars -- as well as their shapes --
to distinguish them from galaxies.
Stars follow a distinctive path in multidimensional
color space, 
and in Section 3, we demonstrate
that their position along this path 
is a function of spectral type.
We then use this stellar locus to assign several new quantities 
to all objects in one of the SDF catalogs.
We show in detail how one can
use these new derived quantities
to select objects more or less likely to be
stars in Section 4.
In Section 5, 
we mention the limitations of our
method, which cannot be applied to the
full depth of the SDF images.
Within its limits, we show that our
sample of stars in the SDF is consistent
with a model of the Milky Way's 
stellar component.

\section{Defining the Stellar Locus}


\cite{Kashikawa2004} 
describe the acquisition of the images which make up the SDF, 
the processing of the images, 
and the creation of a set of catalogs
of objects detected in each passband.
They placed the processed images
and catalogs of objects measured in the field
into a public archive,
the SDF Data Release.
Our analysis is based on these data;
in particular,
we chose to focus on objects matched to detections
in the $R_c$ passband,
since it lies in the middle of the spectral range
covered by the Suprime-Cam optical filters
and since the $R_c$ images are among the
deepest in the group.
The SDF catalogs matched to the $R_c$ passband
contain over $209,000$ objects,
the majority of which are distant galaxies.
In order to study the properties of the stars
which are scattered among these galaxies,
one needs a reliable way to distinguish the
stars from other objects.
Our method requires knowing precisely the region
in color-color space occupied by stars,
so the first step is to define this stellar locus.


We begin by selecting a
``clean'' subset of stars
from this very large catalog.
We tried using the 
{\tt class\_star} values included in 
Public Data Release.
The SExtractor program
(\cite{Bertin1996})
produces this number, which describes the
degree to which an object resembles the PSF;
its value ranges from 0.0 (for very extended objects)
to 1.0 (for perfect point sources).
After examining a number of sources near the
center of the SDF carefully, 
we found that using a simple limit
on this {\tt class\_star} parameter
did a very good, but not perfect, job
of identifying stars.
Even if one placed a very narrow limit on 
the range of acceptable values --
for example, $0.95 < {\tt class\_star} < 1.0$ --
a small number of galaxies would still
contaminate the sample of stars.

Therefore, we searched for an additional
piece of morphological information 
to improve the star/galaxy separation.
\cite{Kashikawa2004} suggest a combination
of magnitude and Full-Width at Half-Maximum (FWHM).
When we used this criterion,
we found that a substantial number of galaxies
were still mixed with the stars.
After some experimentation, we found
another parameter which may do a slightly
better job of identifying stars:
the difference of magnitudes
measured through two different circular apertures.
Using the catalog fields
$m_2$, the magnitude through an aperture
of radius 2 arcseconds, and 
$m_3$, the magnitude through an aperture
of radius 3 arcseconds,
we define the quantity
$\delta \equiv m_2 - m_3$.
The exact value of this quantity for a 
point source will depend upon the size
of the PSF.
Since the SDF images in all passbands
have the same size (a FWHM
of $0{\rlap.}^{''}98$ arcseconds),
we expected that stars would have
roughly the same value of $\delta$ 
in all passbands.
Our inspection of the images and catalogs
showed this to be true:
in all images, isolated stars fell into the range
$0.10 < \delta < 0.18$.

In order to create a ``clean'' set of stars,
we combined the two types of morphological information
in the following manner.
We looked at the information for each
object in the SDF catalogs,
as measured in each of the five wide passbands:
$B, V, R_c, i', z'$.
Each object was given an initial
score of zero.
For every {\tt class\_star} value 
in a passband which fell into the acceptable
range of $0.90 - 1.00$,
we added 1 to its score.
For every $\delta$ value 
in a passband which fell into the acceptable
range $0.10 - 0.18$,
we added 1 to its score.
The maximum score for a truly point-like
object was therefore 10:
5 acceptable values of {\tt class\_star}
and 5 acceptable values of $\delta$.
The minimum score, 
for an object which was obviously extended
in all five passbands,
was 0.

The two methods yield similar results,
as one would expect:
$8570$ objects out of the $209,452$ in the SDF
catalogs
have a {\tt class\_star} value within 
the acceptable range in all five passbands,
and $5431$ objects of the $209,452$
have a $\delta$ value within the acceptable
range in all five passbands.
A total of $3725$ objects, or about $1.8\%$ 
of all detected objects, score a perfect 10,
showing star-like shapes in all five passbands.

\begin{figure}
  \begin{center}
    \FigureFile(80mm,80mm){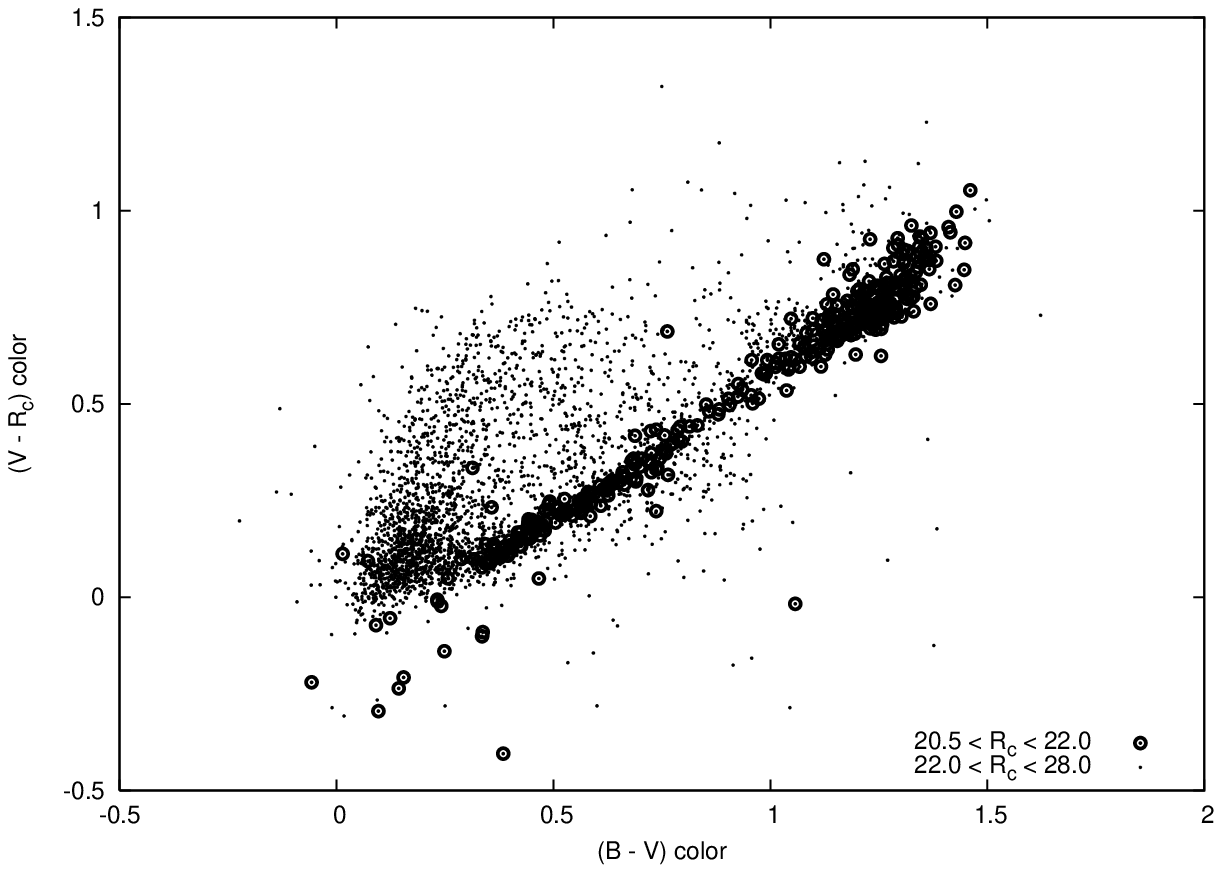}
    \FigureFile(80mm,80mm){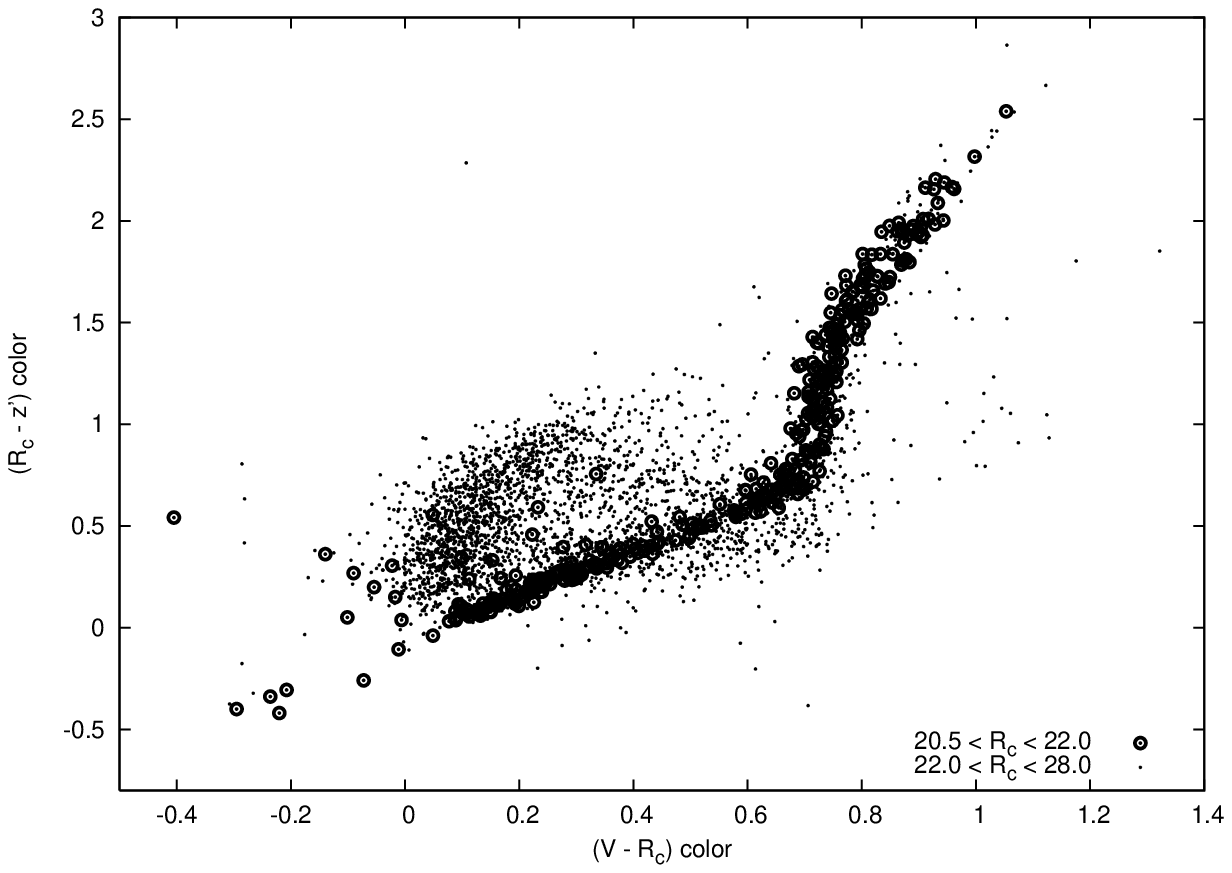}
  \end{center}
  \caption{Colors of objects with morphological score of 10 out of 10: very point-like shapes.}\label{fig:stargalcandidatea}
\end{figure}

However, not all of these objects really are stars.
We know from many previous studies
(e.g. 
{\cite{Johnson1953}}, 
{\cite{Newberg1997}}, 
{\cite{Finlator2000}})
that stars inhabit a rather narrow and
well-defined region in color-color space.
We choose three colors based on the
wide-band {\tt isocor} magnitudes reported in the SDF catalogs:
$(B-V)$, $(V-R_c)$, $(R_c-z')$.
If we plot the locations of the star-like objects
in the space defined by these colors
(see 
Figure \ref{fig:stargalcandidatea}
),
we see two main groups of objects:
a set of mostly bright sources which form a relatively
narrow band, and a set of mostly faint sources
which spread out over a larger region.
Because the brighter sources fall into the region
of color space populated by stars in other studies,
and because galaxies are expected to outnumber stars
greatly at fainter magnitudes,
we focus on the brighter sources to define
our stellar locus.
We created a ``clean'' sample of objects believed
to be stars by choosing objects 
between $20.5 < R_c < 22.0$
with morphological scores 10 out of 10, 
and discarding the few outliers which fell
far from the main group.
Our final sample contains 351 objects.

As a check that our ``clean'' sample really does
consist of stars, 
we examined the proper motions of objects of
objects with morphological scores of 10 out of 10.
The USNO B1.0 catalog 
(\cite{Monet2003})
provides proper motions for a small fraction of
the brightest objects in the SDF.
We determined that spurious matches between the 
catalogs outnumber true matches 
when the matching radius is about 1.0 arcsecond;
within this radius, only about 20 percent of
the matches are spurious.
We therefore matched items in the USNO B1.0 catalog to objects
with morphological scores of 10 out of 10,
using a tolerance of 1.0 arcsecond.
Table {\ref{tab:clean_pm}}
shows the properties of the relatively small number of
objects matched to USNO B1.0 entries.
The majority of matched items have 
proper motions listed as zero;
however, for those with non-zero values,
objects within our ``clean'' subset
clearly have larger motions than those we discarded.
Moreover, the measured proper motions are largest
for the red half of our ``clean'' subset,
as one would expect if the cool stars
were dwarfs.
We conclude that the proper motions provide a
weak piece of
evidence that our ``clean'' sample
is dominated by real stars,
though we cannot be sure without
spectra for each candidate.


Within our three-dimensional color space,
stars inhabit a region which twists and turns.
In order to follow this locus accurately, 
we follow the instructions given by 
{\cite{Newberg1997}}
(see also {\cite{Richards2002}})
to build a three-dimensional 
volume which contains the stars.
These authors note that the stellar locus
can be well approximated by a tube
which is elliptical in cross section
and changes shape and orientation 
as it moves from hot, blue stars 
to cool, red ones.
The basic idea is to build the stellar locus
iteratively:
start by placing a few ``locus nodes''
in the middle of the stellar distribution
(we started with three),
then insert new locus nodes between existing
ones.
The stars surrounding each locus node
are used to define an ellipsoidal
tube which runs between it and its neighbors.
Our code follows the algorithm described
by {\cite{Newberg1997}} very closely;
we added only one feature, a condition on the
minimum number of stars which must belong
to a locus node in order for it to be
allowed to split
(we needed this condition because the number
of stars in our dataset was much smaller
than those in most other situations).

After 100 iterations of the algorithm,
our initial set of 3 locus nodes
grew to 22 locus nodes.
The nature of this technique forces
the nodes at each end of the locus to 
migrate inwards,
leaving a sparse set of objects
exposed at both the blue and red end.
After examining the colors of all objects
in the field surrounding the ends
of the locus,
and considering the synthetic colors
of stars of very early and very late spectral
types (see Section 3),
we extended the locus at each end by adding
one node manually.
Since these nodes were not defined 
by a set of associated objects
following the algorithm,
they did not have a well-defined
ellipsoidal tube.
Therefore, we assigned to each of these
extended nodes a circular tube with
radius equal to the major axis of the nearest
normal locus node.

In Table 
\ref{tab:stellarlocus}, 
we list the nodes which make up our stellar locus.
We provide the location of each node
in color-color space,
the major ($a$) and minor ($b$) axes of the
ellipse defined by star surrounding the node,
and the angle (in radians)
describing the orientation of the major axis of that
ellipse (see 
{\cite{Newberg1997}}
for details of the coordinate system used to 
define this angle).
We also list a quantity we shall call the
``milestone:'' 
the distance along the stellar locus from
locus number 1 to a location along the locus,
measured by moving from locus node to locus node in straight
line segments.
The two end nodes,
numbered 0 and 23, 
are the extensions we added manually.
The columns for $a$ and $b$ show that,
as others have found in different datasets,
stars lie within a relatively narrow
and flattened ribbon in color space:
the ellipse describing the 
stellar locus has a width of less than
$0.08$ magnitudes, and a major axis
usually two or more times larger
than its minor axis.

\section{Stellar Types along the Locus}

The stellar locus we have created
is simply a map of the region in color-color
space in which stars are likely to fall.
It does not provide any information on 
the nature of stars as a function of 
their position along this locus.
However, if we can calculate the location
in this same color space of
either real stars with known properties,
or models of stellar atmospheres,
then we can draw a connection between
the colors of any star in the SDF
and its physical nature: spectral type,
temperature, and even, to a lesser degree,
luminosity or mass.

\begin{figure}
  \begin{center}
    \FigureFile(80mm,80mm){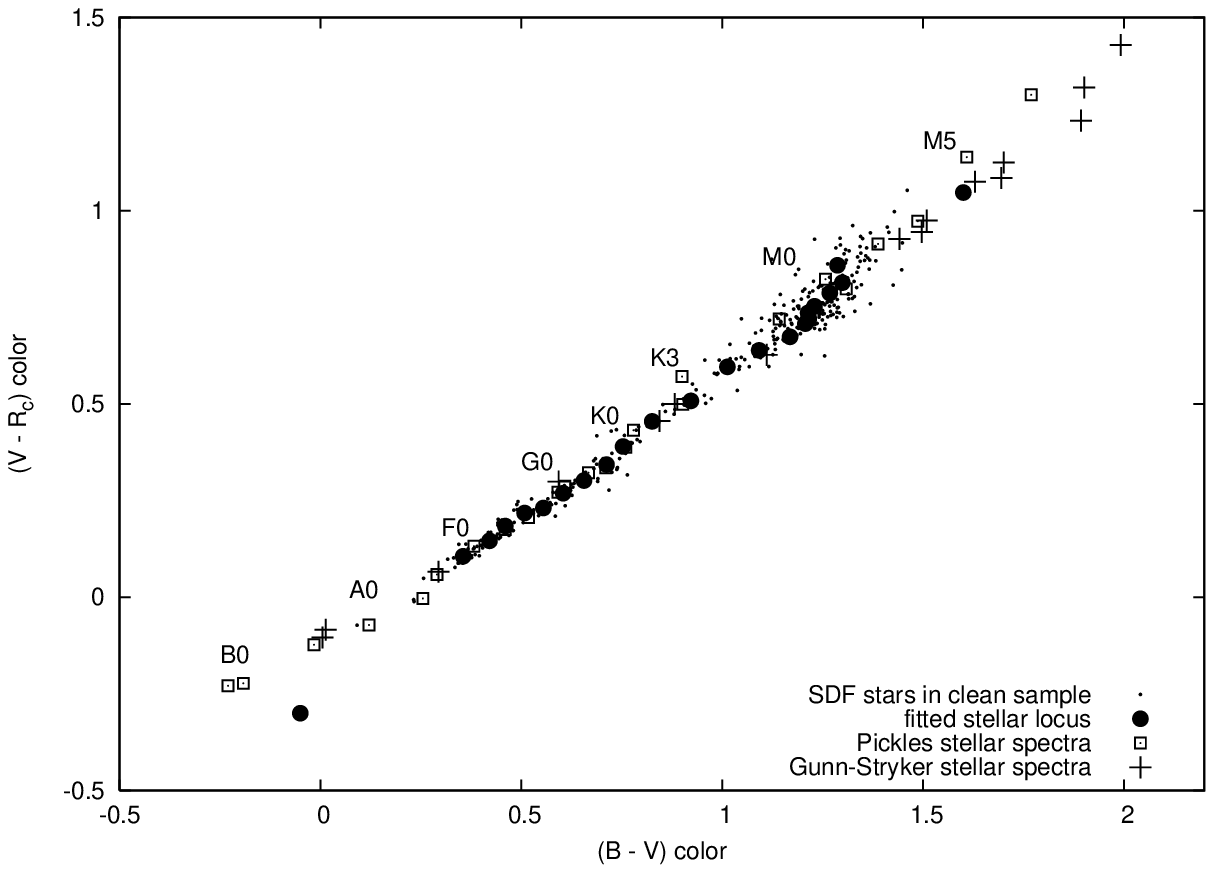}
    \FigureFile(80mm,80mm){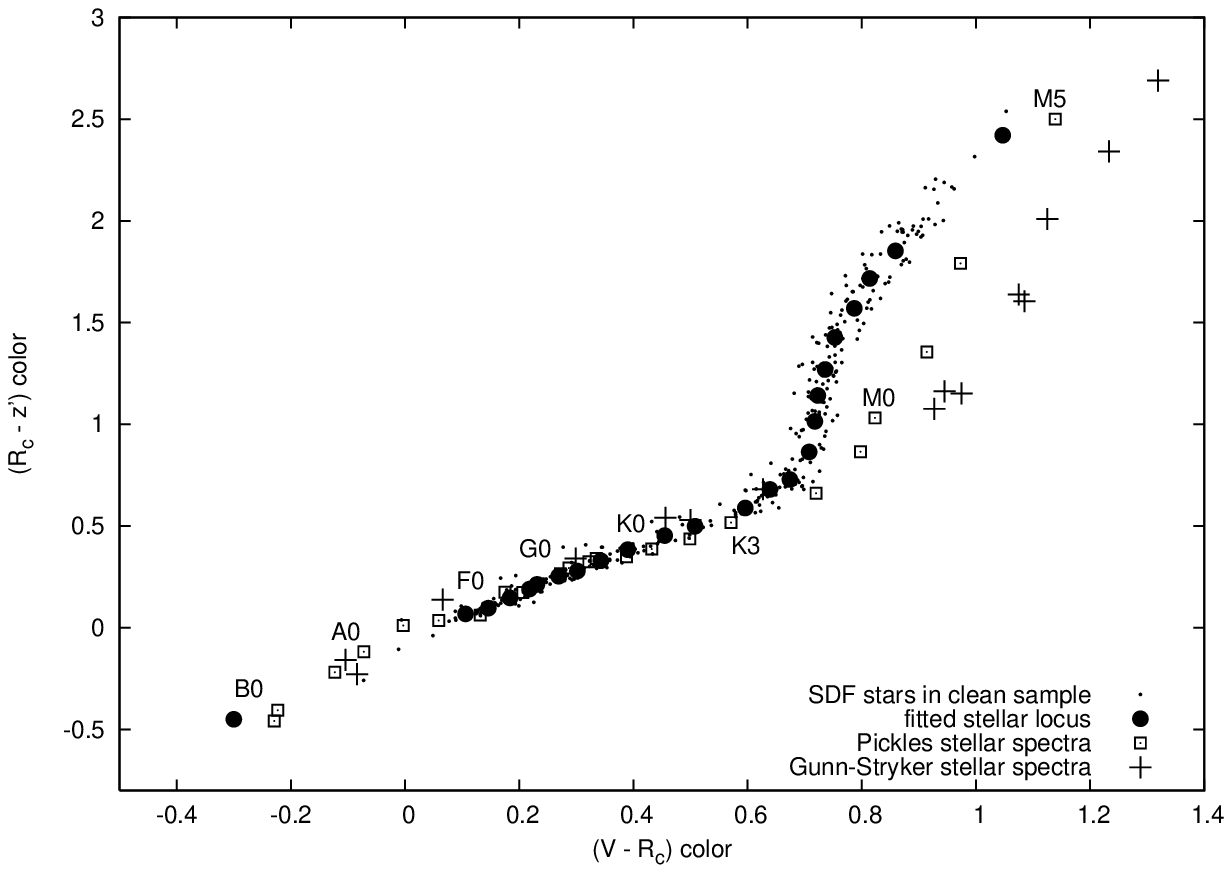}
  \end{center}
  \caption{Observed colors of stars in the SDF compared to synthetic 
             colors of main-sequence stars in spectrophotometric atlases.
             The labels correspond to colors of 
             stars from \cite{Pickles1998}.
   }\label{fig:comparesyn}
\end{figure}

Therefore, we computed
synthetic colors for several sets of
stars with good spectrophotometry and
known spectral class.
We adopted the overall response
curves for Suprime-Cam shown
in Figure 1 of
\cite{Kashikawa2004};
that is, the curves which include the
effects of both instruments and atmosphere.
We then convolved two sets of spectra 
with these passbands to generate synthetic
photometry:
\begin{itemize}
\item{ \cite{Pickles1998}, each entry of which is the composite
                   of several observed spectra }
\item{ \cite{Gunn1983}, each entry of which is the observed spectrum
                   of a single star}
\end{itemize}
Since the zeropoint of the magnitude scales 
are not simply related to the stars found
in the 
\cite{Gunn1983} library,
and since the 
\cite{Pickles1998} data have been normalized
to an arbitrary scale,
we shifted the zeropoints of our synthetic
magnitudes by a small amount so that the
synthetic colors best matched the observed colors.
We compare the observed colors to the synthetic
colors of main sequence stars in Figure 
\ref{fig:comparesyn},
placing labels for each spectral class based on 
main sequence stars from \cite{Pickles1998}.
The synthetic colors lie in the same 
locations as the
observed colors for the most part,
but there are a few regions in which they differ.

At the blue end of the stellar locus,
there are very few observed stars:
the bluest node defined by stars
in our ``clean'' sample corresponds 
roughly to spectral class A7.
In order to extend the locus to hotter
stars, we relied in part on the synthetic colors,
but also gave some weight to the 
observed positions of very blue, 
nearly point-like objects in the
SDF catalogs.
Especially in the graph of $(V - R_c)$ vs. $(B - V)$,
our extended locus strays somewhat
from the synthetic colors.
It would be best, of course, to use observations 
of a field containing very blue stars
to define this end of the locus.

At the cool end of the locus,
we again find very few stars in the
``clean'' sample in the SDF.
This is largely a selection effect,
as explained in Section 5 below.
The last node created by our automatic
algorithm corresponds roughly to 
spectral type M3;
we added one more node to the locus
guided by the colors of the very few
reddest stars measured in the SDF.

The obvious 
discrepancy between the observed colors
and synthetic colors
of cool stars (spectral classes K5 - M5)
is due to differences in metallicity.
The cool main-sequence Pickles and Gunn-Stryker stars
were chosen to be bright enough
($9 < V < 13$) 
to yield spectra with high signal.
Since these cool dwarfs have low luminosities,
they must be within roughly 20 pc of the Sun,
and thus members of the disk population with
high metallicity.
Any M dwarfs observed in the SDF,
on the other hand, 
appear
roughly ten magnitudes fainter
than the spectrophotometric standards.
They are therefore roughly 100 times
more distant, at distances of
several kpc from the Sun and the disk.
We expect them to have lower metallicities.  
As a check, we convolved model stellar spectra
of 
{\cite{Lejeune1997}}
and
{\cite{Lejeune1998}}
with the Suprime-Cam passbands
and computed their colors.
We found that as the metallicity of
cool main-sequence dwarfs decreases 
from 
$[Fe/H] = 0.0$ to $[Fe/H] = -2.0$,
the bend in the stellar locus in
the $(R_c - z')$ vs. $(V - R_c)$
diagram becomes sharper,
matching the observed stellar locus.

Despite this shift at
the cool end,
we believe that the progression of spectral
types along the observed locus 
will mirror the progression of 
spectral types in the spectral libraries.
Therefore, we assign a spectral class to
objects along the observed stellar locus
in the following manner:
for each main spectral class (A0, B0, etc.),
we compute synthetic colors based on
the
\cite{Pickles1998} 
spectrophotometry.
We then find the location along the
observed stellar locus which is closest in
three-dimensional color space to the
theoretical value,
and note the ``milestone'' 
(distance along the locus measured
from node 1)
at that location.
Table 
\ref{tab:spectralclass} 
lists the relationship between 
spectral class and 
a measured property of objects in the 
SDF.
Once we have found the position of an
observed star in the SDF along the 
stellar locus,
we can use its
``milestone''
to estimate a rough spectral class.

We remind the reader that the bright, point-like objects
in the SDF which we use to construct the stellar locus
span a somewhat limited range of spectral types:
from about A7 at the blue end to about
M3 at the red end.
Using our locus to assign spectral class to stars
outside this range is less reliable.

\section{Using the Stellar Locus to Identify Stars}

In Section 2,
we used a small set of relatively bright, point-like
objects -- the ``clean'' sample of 351 stars --
to build a stellar locus in color space.
We can now apply this locus as a tool to 
search for additional stars in the total 
set of objects in the SDF catalogs.
We describe in this section a method which gives
each object a ``color-based'' probability of being
a star,
which is independent of the ``shape-based'' probability
described earlier.
A combination of both methods yields a more reliable
sample of stars than either method alone.

The procedure for computing the ``color score''
of an object begins
with its location in the three-dimensional
color space of 
$(B-V)$, $(V-R_c)$, $(R_c - z')$.
We find the position along the stellar
locus which is closest to the star's color.
We then ask the question: is the distance
of this star from the stellar locus inside
or outside the ellipsoidal tube defined
by stars surrounding the locus at this location?
The 
{\cite{Newberg1997}}
algorithm determines the elliptical
shape of the tube at each locus node,
defined in a plane which is perpendicular
to the line segment joining one locus
node to the next.
Following their example,
we extend the ellipse outwards to its
$3\sigma$ boundary;
in other words, we make the extent of the
tube large enough that it contains
(on average) $98\%$ of the stars in the local
region of color space.
Now, in general, a stellar candidate
will be closest to some location
between two locus nodes: call them
$P_i$ and $P_{i+1}$.
In some sections of the stellar locus,
such as the stretch from spectral class F0
to spectral class G0, 
the locus lies along a straight line in color
space, and the elliptical shape remains
relatively constant.
But in other regions, such as the 
stretch from spectral class K0 to spectral class M0,
the locus makes sharp bends and twists.
If the shape and orientation of the tube at $P_i$ are
somewhat different than the shape and orientation
at $P_{i+1}$, which elliptical parameters should we use when
evaluating the nature of a candidate between them?

\begin{figure}
  \begin{center}
    \FigureFile(80mm,80mm){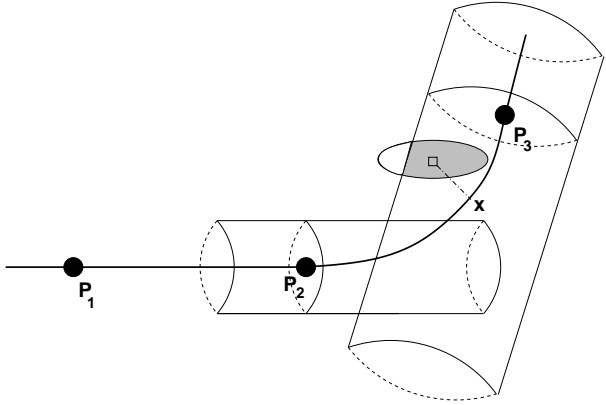}
  \end{center}
  \caption{Giving an object its ``color score:'' the observed colors
             of an object (square) place its closest approach to the 
             stellar locus, at $x$, between locus nodes $P_2$ and $P_3$.
             The cloud of test positions around the object lies
             partially within the tube extending backwards from $P_3$.
             This object will receive a ``color score'' of about 0.75.
   }\label{fig:tubes}
\end{figure}

We take an inclusive approach:
we extend the tube forward from locus node $P_i$
using its parameters
and determine whether the candidate lies inside
or outside that tube;
and we also extend the tube backwards from locus node $P_{i+1}$
using {\it its} parameters
and determine whether the candidate lies
inside or outside that tube.
If the candidate lies within either tube,
then we consider it consistent with the stellar locus.
A candidate must lie outside both tubes to fail the test.
The candidate shown by a square in Figure 
\ref{fig:tubes}
falls outside the tube extended forward from locus
node $P_2$, but inside the tube extended backwards from $P_3$.

There is an additional complication:
the magnitudes for every object in the SDF catalogs
contain a tabulated uncertainty,
which means that the colors derived from those
magnitudes also have associated uncertainty.
We use a Monte Carlo approach to assign a 
score to each object in the following way:
for each real observed object,
we generate $N = 100$ 
test points
by adding random gaussian errors to
the measured magnitudes $B$, $V$, $R_c$, $z'$.
These test points appear as a cloud in color space
surrounding the actual measured position of an object.
Following the method described above,
we determine if each of these test points
lies inside or outside the stellar locus.
We then assign a ``color score'' to the object
which is simply the fraction of all test points
which fall within the stellar locus.
For example, roughly three-quarters of the 
test points around the object in Figure 
\ref{fig:tubes}
fall within the tube extending backwards
from locus node $P_3$;
therefore, this object would receive a ``color score''
of about 0.75.

We have applied this procedure to all
the objects in the 
SDF $R_c$-based catalog,
regardless of their shape or size.
A summary of the results is shown in 
Table 
\ref{tab:starsummary}.
It is clear that color provides a 
sharper tool than shape to divide stars from 
galaxies.
For example, roughly 12\% of all sources
have shapes in the different images
which are at least somewhat point-like
(scoring at least 5 out of 10 on the
morphological scale),
but fewer than 2\% of all sources
have colors which are somewhat star-like
(scoring at least 0.5 out of 1.0 on
the color scale).


\section{Discussion}

Before we can interpret the results of our
classification, we need to account for the 
selection effects in its construction.
The original SDF observations had a range of 
total exposure times, 
from 340 minutes in $V$-band to 801 minutes in
$i'$-band.
The resulting images reach different
depths.
We have concentrated on the catalogs produced
based on detections in the $R_c$ images,
and requiring matching detections in the other passbands.
If we examine the distribution of objects
as a function of magnitude in each passband,
we find a similar pattern in each case:
a few very bright objects, then increasing numbers
of fainter objects, reaching a peak at some
particular magnitude and then declining.
The peak of this distribution is fainter
for fuzzy sources than for point-like sources
in $B$, $V$, $R_c$, but roughly the same for $z'$.
It is important that the peak magnitude is 
not the same in all passbands: 
we find the peak for point-like objects 
to be between $25.4$ and $25.8$ for
$B$, $V$ and $R_c$, but only $24.4$ for $z'$;
we will refer to these values as the
``multicolor completeness limit'' for point sources.


This creates a significant bias against faint
objects of extreme colors
in the catalogs.
Very blue objects (say, stars of spectral
type B8) which are close to the plate
limit in the $R_c$ band will fail to appear in the
$z'$ images and so not be included in the 
catalogs.
Very red objects (say, stars of spectral
type M3), on the other hand,
which are barely detected in the $z'$ images
will not appear at all in the $B$ or $V$ images,
and so again fall out of the catalogs.
In order to compare our results to those
in other regions of the sky, 
we need to include the effects of these selection biases.

\begin{figure}
  \begin{center}
    \FigureFile(80mm,80mm){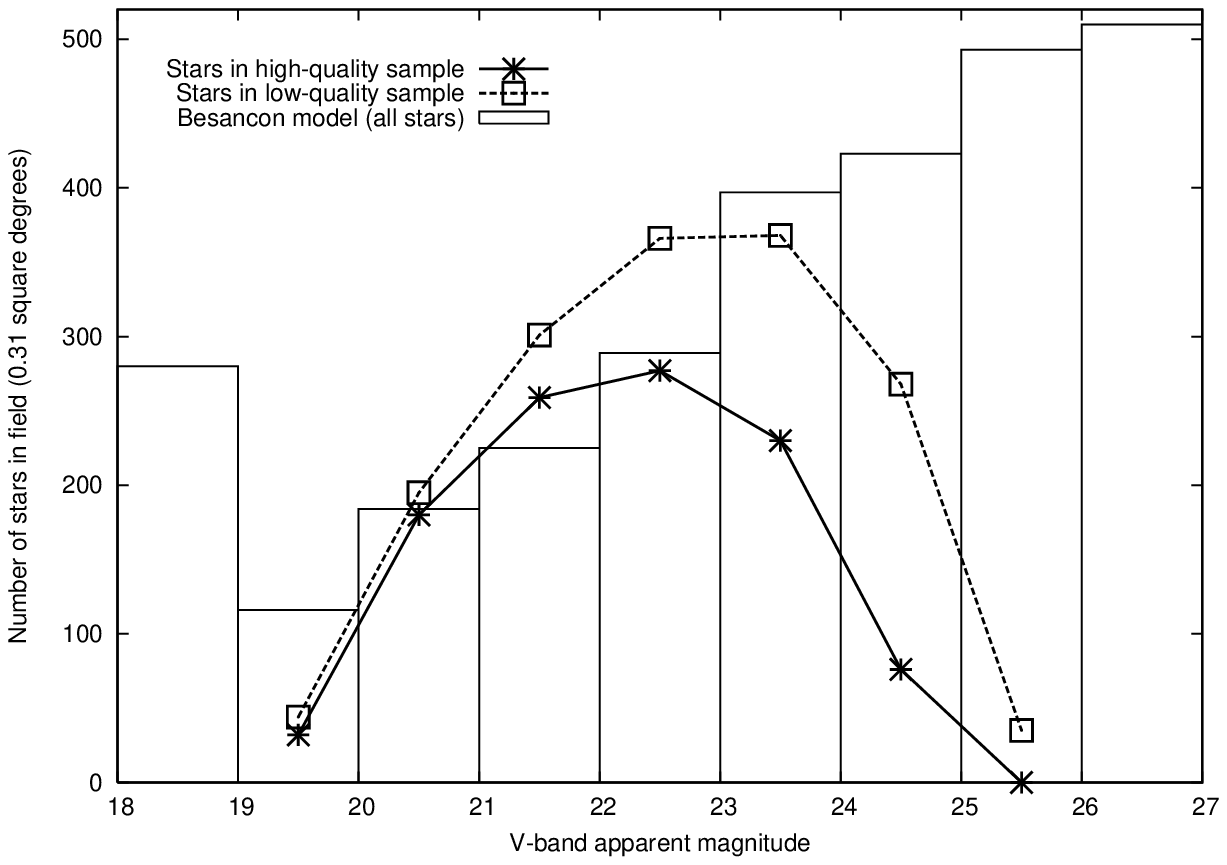}
    \FigureFile(80mm,80mm){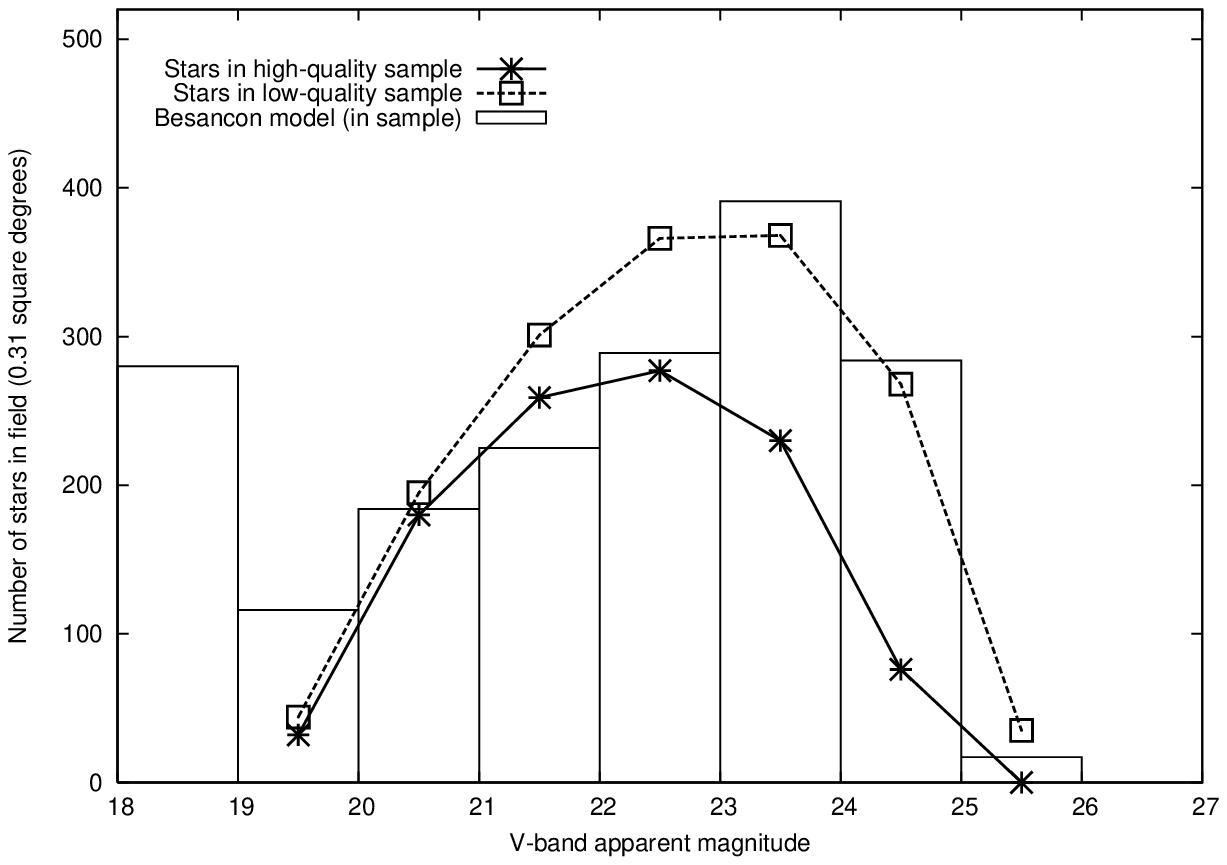}
  \end{center}
  \caption{Counts of objects classified as stars by our algorithm and stars in the Besan\c{c}on model of the field, as a function of $V$-band magnitude.  In the right-hand panel, we have discarded stars from the model which would not appear in the input to our algorithm.
   }\label{fig:modelcounts}
\end{figure}

We have found the galactic population model
produced by astronomers 
at the Besan\c{c}on Observatory
(\cite{Bienayme1987}, \cite{Haywood1997}, \cite{Robin2003}; 
hereafter ``the Besan\c{c}on model'')
very useful,
especially since it is easy to run 
interactively: see
{\tt www.obs-besancon.fr/modele/modele.html}.
We generated a simulation of stars that should
appear within the region of the SDF,
using the default Besan\c{c}on model parameters
and a maximum $V$-band observed magnitude of $28.0$.
If we compare the simulation to the objects
classified as stars by our algorithm
(see Figure
\ref{fig:modelcounts}
),
we see that our catalog follows the model
only until magnitude $V \sim 23$.
However, if we generate magnitudes in the SDF passbands
for each star in the model, 
and discard any star which falls below the 
multicolor completeness limit 
for point sources in our input catalogs
($B = 25.8$, $V = 25.6$, $R_c = 25.4$, $z' = 24.4$),
then we find the number of stars in the
model matches the number of objects we classify
as stars reasonably well.


  

The good agreement between 
our sample of objects classified as stars
and 
the Besan\c{c}on model 
suggests that we may use the 
model to provide insight into some
properties of our sample.
For example,
we note that of the $1786$ stars in the
Besan\c{c}on model which would qualify
as stars in our algorithm,
only $36$, or roughly $2\%$, 
have evolved far off the main sequence.
If the great majority of stars
we observe are on the main sequence,
then we would not expect to see any
significant split of the stellar locus into 
two tracks for giants and dwarfs.
As described above in Section 3,
we indeed do not see any indication of 
a bifurcation in the stellar locus.
Moreover, we are apparently
justified in using spectra of 
main sequence stars to compute synthetic
colors in the Suprime-Cam system.
The Besan\c{c}on model
also predicts that many of the 
cool stars in our observed sample
ought to have low metallicities,
which again agrees
with the observations.

As an additional check on our separation of stars
from galaxies,
we considered proper motions.
Using a 1-arcsecond radius,
we found 1605 matches between the SDF catalog
and the USNO B1.0;
253 of these matched objects fell into
our ``low-quality star'' category,
while the remaining 1352 were classified
as non-stellar.
A majority ($58\%$) of objects 
classified as stellar had a non-zero proper motion,
while a smaller fraction ($37\%$)
of the objects classified as non-stellar had
a non-zero proper motion.
At the current time, this kinematic evidence 
only weakly supports our claim to identify
stars.  We hope to revisit the issue
after future observations provide more
precise measurements of proper motions.


\begin{figure}
  \begin{center}
    \FigureFile(80mm,80mm){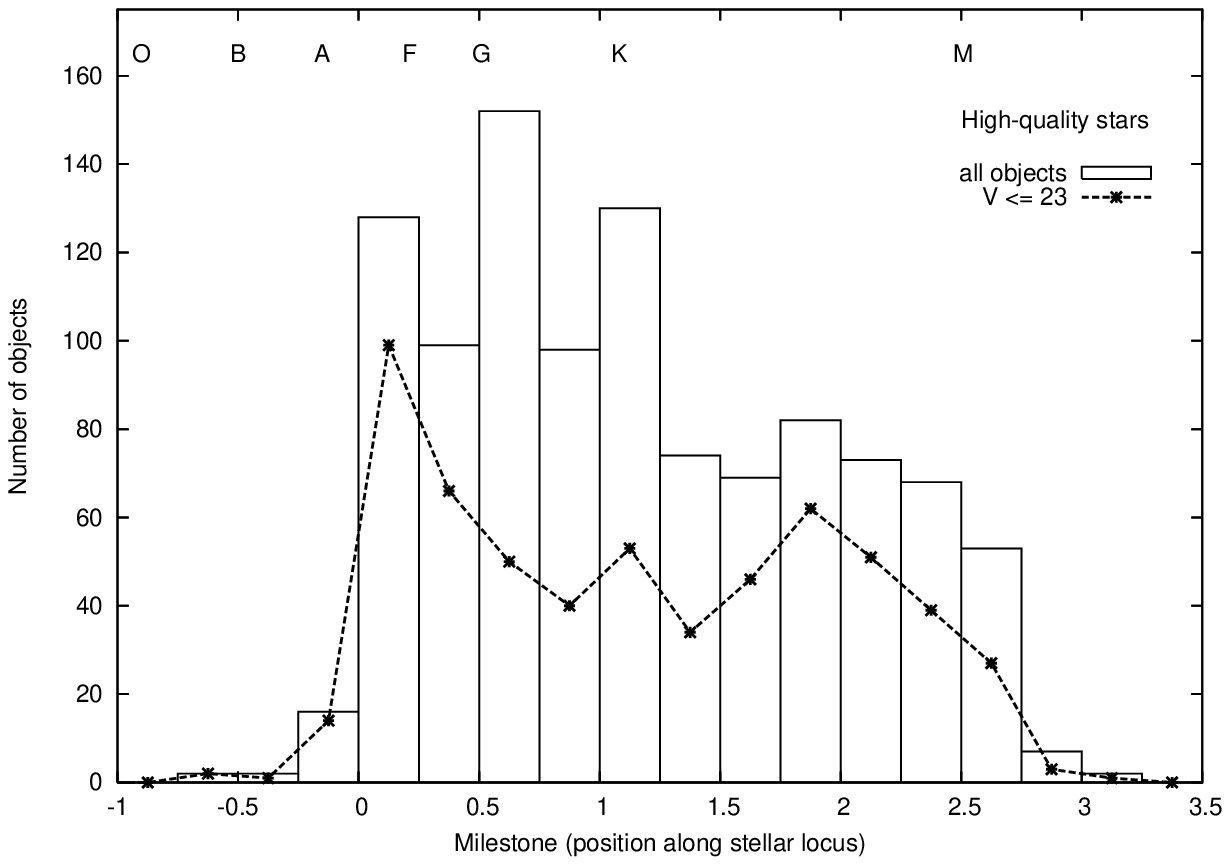}
    \FigureFile(80mm,80mm){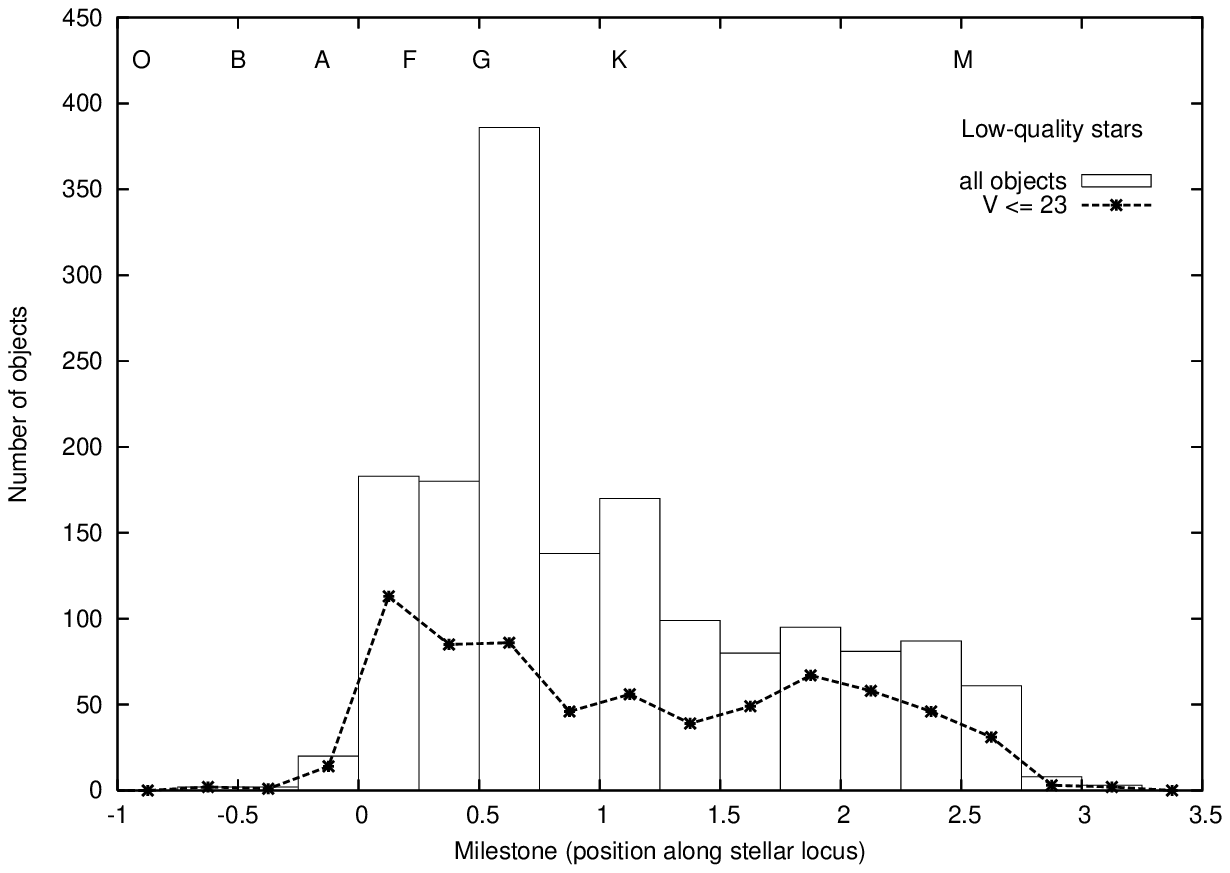}
  \end{center}
  \caption{Distribution of the ``milestone'' values (location along the stellar locus) of objects classified as stars.  The labels at top are in the middle of each spectral type's range.
   }\label{fig:milestones}
\end{figure}

Our algorithm not only separates stars
from galaxies based on their colors,
but also provides rough spectral classifications,
as shown in Table
{\ref{tab:spectralclass}}.
We show the distribution of 
``milestone'' values for the objects
classified as stars 
in Figure
{\ref{fig:milestones}}.
Both the high-quality and low-quality 
samples show a similar pattern:
the distribution of the brightest stars is somewhat bimodal,
the majority located at either the blue end 
(spectral type A-F) or red end (spectral type M)
of the stellar locus.
Other photometric studies 
of our galaxy at these magnitudes
({\cite{Infante1986}, {\cite{Infante1994}, {\cite{Prandoni1999})
show this double-peaked distribution.
The standard explanation 
(e.g., {\cite{Bahcall1986}})
is that the
blue peak indicates the main-sequence turnoff
point of metal-poor halo stars,
while the red peak is due to the competition
between increasing number and decreasing luminosity
of low-mass disk stars.

In both samples, however, 
when we look 
fainter than the multicolor completeness
limit,
we see a large number of objects with 
intermediate colors.
The excess is especially obvious in the
low-quality sample.
We examined these faint, intermediate-color
objects carefully to see if there might be
some obvious explanation.
They do not show any clumping or patterns
in their (RA, Dec) positions in the field,
so they are not artifacts of diffraction spikes
or halos around bright objects.
A significant fraction, perhaps one-third,
of the items which appear only in the low-quality
sample are blends of two or more objects;
it makes some sense that blended objects would
have typical colors in between those of 
individual objects.
Some of these sources appear slightly elliptical,
and may be nearly round galaxies.
However, we also find that many of these
objects appear to be reasonably isolated and
point-like.
Could there really be an increase 
in the number of stars of intermediate color
at magnitudes $V > 23$?
We remind the reader that this is below our
multicolor completeness limit, so that many stars
at both the blue and red ends of the population 
are not included in the sample.
We also note that a main-sequence G star
of absolute magnitude $M_V \sim 5$ and
apparent magnitude $m_V \sim 24$ would be
located at a distance of roughly
$d \sim 60$ kpc, 
very far from both the plane and the
bulge of the Milky Way Galaxy.
The Besan\c{c}on model does not show this
feature.
If this feature is real, 
we speculate that it might be due to a local
perturbation to the normal halo population,
perhaps the remnant of some disrupted satellite galaxy
or globular cluster.
According to 
\cite{Martinez2001}
and
\cite{Majewski2003},
the SDF lies within ten degrees
of the northarm arm of the Sagittarius Dwarf stream,
which has a distance of roughly 50 kpc from the Sun
at this location.


\section{Conclusions}

We have used one set of the Subaru SDF 
catalogs, based on detections in the 
$R_c$ images and corresponding measurements in 
in $B$, $V$, $R_c$ and $z'$,
to study stars in the SDF.
We used both morphological and color 
information to choose a small set of 
certain stars,
around which we created
a stellar locus in three-dimensional
color space.
We then compared the location in color-space
of {\it all} sources in the SDF catalogs to
the stellar locus.
The results provide a quantitative
measure which not only distinguishes stars from
galaxies, but also 
assigns rough spectral types to the stellar
candidates.
We find good agreement between
the Bescan\c{c}on model of the Milky Way
and 
number counts of stars in the SDF
down to our ``multicolor completeness limit''
(which is several magnitudes brighter
than the detection limit in any 
individual passband).
Our catalog of objects in the SDF,
which includes their proximity to
and position along the stellar locus,
can be found at
{\tt http://spiff.rit.edu/sdf/}.

\bigskip
We thank the referee, Heidi Newberg, for suggestions
which improved the paper.


\begin{longtable}{lccrrr}
  \caption{Proper motion ($PM$), in units mas/yr, of objects with morphological scores 10 out of 10}\label{tab:clean_pm}
  \hline\hline
 Subset & Number & $PM > 0$ & \multicolumn{3}{c}{For objects with $PM > 0$}  \\
        &                  &          & mean & stdev & median \\
\endfirsthead
  \hline\hline
\endhead
  \hline
\endfoot
  \hline
\endlastfoot

   in clean set, $(V-R_c) < 0.6$ & 64 & 20\% & 57 & 68 & 37 \\
   in clean set, $(V-R_c) > 0.6$ & 58 & 31\% & 74 & 60 & 43 \\
   outside clean set  & 36 & 22\% & 12 & 4 & 12 \\
\end{longtable}

\begin{longtable}{llllllll}
  \caption{The stellar locus.}\label{tab:stellarlocus}

  \hline\hline
 Node & $(B - V)$ & $(V - R_c)$ & $(R_c - z')$ & Milestone & \qquad $a$ & \qquad $b$ & \ \ $\theta$  \\
  
\endfirsthead
  \hline\hline
\endhead
  \hline
\endfoot
  \hline
\endlastfoot

  0  & -0.050 & -0.300 & -0.450  &   -0.773  & 0.0258  & 0.0258   & 2.85    \\
  1  & \phantom{-}0.355 & \phantom{-}0.106 & \phantom{-}0.068  &    \phantom{-}0.000  & 0.0258  & 0.0122   & 2.85    \\
  2  & \phantom{-}0.421 & \phantom{-}0.146 & \phantom{-}0.096  &    \phantom{-}0.082  & 0.0145  & 0.0077   & 2.79    \\
  3  & \phantom{-}0.460 & \phantom{-}0.184 & \phantom{-}0.147  &    \phantom{-}0.156  & 0.0375  & 0.0163   & 3.02    \\
  4  & \phantom{-}0.508 & \phantom{-}0.218 & \phantom{-}0.190  &    \phantom{-}0.229  & 0.0310  & 0.0156   & 2.90    \\
  5  & \phantom{-}0.555 & \phantom{-}0.231 & \phantom{-}0.213  &    \phantom{-}0.283  & 0.0282  & 0.0120   & 2.65    \\
  6  & \phantom{-}0.604 & \phantom{-}0.269 & \phantom{-}0.253  &    \phantom{-}0.356  & 0.0204  & 0.0132   & 2.92    \\
  7  & \phantom{-}0.656 & \phantom{-}0.302 & \phantom{-}0.279  &    \phantom{-}0.424  & 0.0220  & 0.0122   & 3.09    \\
  8  & \phantom{-}0.712 & \phantom{-}0.343 & \phantom{-}0.329  &    \phantom{-}0.509  & 0.0356  & 0.0239   & 0.18    \\
  9  & \phantom{-}0.753 & \phantom{-}0.390 & \phantom{-}0.384  &    \phantom{-}0.592  & 0.0342  & 0.0263   & 1.32    \\
 10  & \phantom{-}0.826 & \phantom{-}0.455 & \phantom{-}0.453  &    \phantom{-}0.712  & 0.0463  & 0.0116   & 3.02    \\
 11  & \phantom{-}0.922 & \phantom{-}0.508 & \phantom{-}0.499  &    \phantom{-}0.831  & 0.0327  & 0.0148   & 2.89    \\
 12  & \phantom{-}1.013 & \phantom{-}0.596 & \phantom{-}0.589  &    \phantom{-}0.986  & 0.0390  & 0.0217   & 2.62    \\
 13  & \phantom{-}1.092 & \phantom{-}0.639 & \phantom{-}0.680  &    \phantom{-}1.114  & 0.0431  & 0.0208   & 2.92    \\
 14  & \phantom{-}1.169 & \phantom{-}0.674 & \phantom{-}0.729  &    \phantom{-}1.212  & 0.0529  & 0.0259   & 2.81    \\
 15  & \phantom{-}1.207 & \phantom{-}0.708 & \phantom{-}0.864  &    \phantom{-}1.356  & 0.0573  & 0.0134   & 2.50    \\
 16  & \phantom{-}1.215 & \phantom{-}0.718 & \phantom{-}1.015  &    \phantom{-}1.507  & 0.0570  & 0.0167   & 2.35    \\
 17  & \phantom{-}1.215 & \phantom{-}0.723 & \phantom{-}1.142  &    \phantom{-}1.634  & 0.0394  & 0.0163   & 1.73    \\
 18  & \phantom{-}1.214 & \phantom{-}0.736 & \phantom{-}1.270  &    \phantom{-}1.763  & 0.0546  & 0.0180   & 2.25    \\
 19  & \phantom{-}1.230 & \phantom{-}0.753 & \phantom{-}1.427  &    \phantom{-}1.922  & 0.0575  & 0.0185   & 2.49    \\
 20  & \phantom{-}1.268 & \phantom{-}0.787 & \phantom{-}1.570  &    \phantom{-}2.074  & 0.0566  & 0.0176   & 2.68    \\
 21  & \phantom{-}1.299 & \phantom{-}0.814 & \phantom{-}1.717  &    \phantom{-}2.227  & 0.0673  & 0.0215   & 1.99    \\
 22  & \phantom{-}1.287 & \phantom{-}0.859 & \phantom{-}1.853  &    \phantom{-}2.388  & 0.0717  & 0.0291   & 1.51    \\
 23  & \phantom{-}1.600 & \phantom{-}1.047 & \phantom{-}2.421  &    \phantom{-}3.063  & 0.0717  & 0.0717   & 1.51    \\

\end{longtable}

\begin{longtable}{ccc}
  \caption{Conversion from position along stellar locus to spectral class.}\label{tab:spectralclass}

  \hline\hline
 Start milestone & End milestone & Spectral class  \\
  
\endfirsthead
  \hline\hline
\endhead
  \hline
\endfoot
  \hline
\endlastfoot

  -1.00  & -0.77  &  O \\
  -0.77  & -0.34  &  B \\
  -0.34  & \phantom{-}0.03  &  A \\
  \phantom{-}0.03  & \phantom{-}0.38  &  F \\
  \phantom{-}0.38  & \phantom{-}0.63  &  G \\
  \phantom{-}0.63  & \phantom{-}1.53  &  K \\
  \phantom{-}1.53  & \phantom{-}3.50  &  M \\
\end{longtable}

\begin{longtable}{ccrl}
  \caption{Summary of classification of 189,380 sources in SDF $R_c$-band catalog.}\label{tab:starsummary}
  \hline\hline
 Shape score (0 - 10) & Color score (0.0 - 1.0)& Number of objects & Category name \\
  
\endfirsthead
  \hline\hline
\endhead
  \hline
\endfoot
  \hline
\endlastfoot

   $0 -  1\phantom{0}$ &  $0.0 - 1.0$ & 127,359 \qquad\qquad &  non-stellar shape \\
   $0 - 10$ &  $0.0 - 0.1$ & 185,819 \qquad\qquad &  non-stellar colors \\
   $5 - 10$ &  $0.0 - 1.0$ & 22,157 \qquad\qquad & somewhat stellar shape \\
   $9 - 10$ &  $0.0 - 1.0$ & 6348 \qquad\qquad & very stellar shape \\
   $0 - 10$ &  $0.5 - 1.0$ & 2711 \qquad\qquad &  somewhat stellar colors \\
   $0 - 10$ &  $0.9 - 1.0$ & 2024 \qquad\qquad &  very stellar colors \\
   $5 - 10$ &  $0.5 - 1.0$ & 1595 \qquad\qquad & low-quality stars \\
   $9 - 10$ &  $0.9 - 1.0$ & 1061 \qquad\qquad & high-quality stars \\
\end{longtable}



\end{document}